\documentclass[pre,twocolumn,showpacs]{revtex4}
\usepackage{epsfig}

\begin{document}

\title{Internal waves in a compressible two-layer atmospheric model:
The Hamiltonian description}
\author{V. P. Ruban}
\email{ruban@itp.ac.ru}
\affiliation{Landau Institute for Theoretical Physics,
2 Kosygin Street, 119334 Moscow, Russia} 

\date{\today}

\begin{abstract}
Slow flows of an ideal compressible fluid (gas) in the gravity field in the
presence of two isentropic layers are considered, with a small difference 
of specific entropy between them. Assuming irrotational flows in each layer
[that is ${\bf v}_{1,2}=\nabla\varphi_{1,2}$], and neglecting acoustic degrees
of freedom by means of the conditions
$\mbox{div}(\bar\rho(z)\nabla\varphi_{1,2})\approx0$, where $\bar\rho(z)$ 
is a mean equilibrium density, we derive equations of motion 
for the interface in terms of the boundary shape $z=\eta(x,y,t)$ and the
difference of the two boundary values of the velocity potentials:
$\psi(x,y,t)=\psi_1-\psi_2$. A Hamiltonian structure of the obtained equations
is proved, which is determined by the Lagrangian of the form
${\cal L}=\int \bar\rho(\eta)\eta_t\psi \,dx dy -{\cal H}\{\eta,\psi\}$.
The idealized system under consideration is the most simple theoretical model
for studying internal waves in a sharply stratified atmosphere, where
the decrease of equilibrium gas density with the altitude due to 
compressibility is essentially taken into account. For planar flows, a
generalization is made to the case when in each layer there is a constant
potential vorticity. Investigated in more details is the system with a model 
density profile $\bar\rho(z)\propto \exp(-2\alpha z)$, for which the Hamiltonian 
${\cal H}\{\eta,\psi\}$ can be expressed explicitly. A long-wave 
regime is considered, and an approximate weakly nonlinear equation of the form
$u_t+auu_x-b[-\hat\partial_x^2+\alpha^2]^{1/2}u_x=0$ (known as Smith's equation) 
is derived for evolution of a unidirectional wave.
\end{abstract}

\pacs{47.10.Df, 47.55.-t, 92.60.-e}

\maketitle

\section{Introduction}

Internal waves constitute an important part in the dynamics of such complex systems
as are Atmosphere and Ocean (see, e.g., Refs.
\cite{C_1989,RE_1993,WSD,KTA_2000,L_T_2001,GPP_2002,VBR_2000,VH_2002,VS_2006,GPT_2007},
and references therein).
These waves are known to propagate at the background of some inhomogeneity
of internal properties of the fluid (gas). In the ocean the main role is played by
salt concentration and temperature, while in the atmosphere the most important
factors are specific entropy and air moisture. Non-uniformity of shear flows
should be mentioned as well. The internal wave dynamics depends essentially on
the condition if the stratification is smooth enough in a wide range of altitudes,
or the change of internal properties takes place sharply near some surface.
The last case, as a rule, is more convenient for a theoretical study, since
the spatial dimensionality of the problem is reduced. 
In many works therefore simplified atmospheric and oceanic models are considered,
where the system consists of several layers, 
with homogeneous fluid within each layer, and then the dynamics of
interfaces between the layers is investigated (see, e.g., \cite{Choi_Camassa_1999,CGK_2005,deZarate_Nachbin,BLS_2008,R_Y_2007,R_2008,Goncharov}, 
and references therein). To the best author's knowledge, in all previous 
finite-layer models the fluid was assumed to be incompressible, even when the
atmosphere was modeled. In the present work, perhaps for the first time, 
an essentially compressible two-layer atmospheric model is considered.
Here it is assumed that there is a sharp boundary $z=\eta(x,y,t)$ separating
two regions of potential flow, with a constant value of specific entropy in 
each layer. The relative difference of that values is small, and it ensures
the slowness of typical flow velocities compared to the local speed of sound.
Accordingly, the acoustic degrees of freedom can be effectively ``filtered'' 
by the conditions $\nabla\cdot(\bar\rho(z){\bf v})=0$ in each layer 
(where $\bar\rho(z)$ is the equilibrium density), instead dealing with the full
continuity equation $\rho_t+\nabla\cdot(\rho{\bf v})=0$.
This idea to eliminate relatively fast sound waves was used previously 
to obtain simplified equations describing convection and internal waves in a
continuously stratified compressible fluid  \cite{Ogura_Phillips,Durran,Bannon}, 
and also slow isentropic vortex flows in a compressible fluid placed in a static
external field \cite{R2000PRD,R2001PRE}. The distinction of the present model
is that the potentiality condition in each layer, together with the equation
$\nabla\cdot(\bar\rho(z)\nabla\varphi)=0$ for the velocity potential, allow
us to represent equations of motion in terms of the interface shape $z=\eta(x,y,t)$ 
itself and the difference of the two boundary values of the velocity
potential. Moreover, we succeeded in proving a Hamiltonian structure of the obtained
equations, which is a generalization of the canonical  structure discovered by 
V. E. Zakharov in the dynamics of waves at the free surface of an ideal 
incompressible fluid \cite{Z68,Z99,ZK97}. In the two-dimensional (2D) case, 
it is possible to consider in the framework of the two-layer model also shear 
flows with piecewise constant potential vorticity. 
The Hamiltonian theory is naturally modified in that case. 
As applications of the developed theory, we obtained the dispersion
relation for internal waves in the two-layer compressible atmosphere, 
and we derived a nonlinear equation which is intermediate between 
the Korteweg-de Vries and the Benjamin-Ono equations \cite{BENJAMIN,ONO}. 
This equation determines slow evolution of a unidirectional wave and 
it takes into account the dispersive correction of a special form, 
taking place in the model. Previously, a similar equation was derived in 
a different physical context by Ronald Smith \cite{Smith_1972}, who investigated
continental-shelf waves in ocean.

The paper is organized in the following way. In Section 2,  simplified
equations for the two-layer compressible atmospheric model are suggested, 
and their Hamiltonian structure is proved. In Section 3, calculations are
performed for the case of exponential profile of equilibrium density, including
derivation of the dispersion relation and the approximate nonlinear equation for
evolution of propagating wave. In Section 4, the generalization of the model to 
2D flows with a piecewise constant potential vorticity is made. 
In Section 5, some conditions of applicability of the model 
are briefly discussed, as well as perspectives of future research. 
Finally, in the Appendix we derive three-dimensional (3D) Green's function
determining the Hamiltonian of the system in the presence of the lower flat
boundary.

\section{Approximate equations and their Hamiltonian structure}

Let us assume that in the equilibrium state the first layer of gas occupies
the region $0<z<h$ and has the density $\bar\rho_1(z)$, while the second layer
occupies the region $z>h$ and has the density $\bar\rho_2(z)$ [for simplicity,
we have supposed that the lower rigid boundary --- ``the Earth surface'' ---
is flat, but the more general case of nontrivial topography can be considered
in analogous way]. Of course, functions $\bar\rho_1(z)$ and $\bar\rho_2(z)$ cannot
be arbitrary, since in fact they are specified by the hydrostatic balance condition
together with an equation of state  of the gas (see below).
For the further derivation of approximate equations describing
potential flows in this system which are slow compared with a local speed 
of sound $c$, the following condition is very important:
$(\bar\rho_1-\bar\rho_2)\ll\bar\rho(z)=(\bar\rho_1+\bar\rho_2)/2$.

The starting-point equations for potential isentropic gas flow in each layer are
the non-stationary Bernoulli equation and the continuity equation,
\begin{eqnarray}
&&\partial_t\varphi+\frac{(\nabla\varphi)^2}{2}=-w(\rho)-gz+const,\\
&&\partial_t\rho+\nabla\cdot(\rho\nabla\varphi)=0,
\end{eqnarray}
where $\varphi({\bf r},t)$ is the potential for the velocity field ${\bf v}$,
satisfying the condition of zero normal derivative at the rigid boundary, that is
$\partial_z\varphi(x,y,0)=0$; $\rho({\bf r},t)$ is the density, 
$w(\rho)$  is the specific enthalpy which is defined by the formula
\begin{equation}
w(\rho)=w_{1,2}(\rho)=\int_0^\rho \frac{dp_{1,2}(\rho)}{\rho}.
\end{equation}
Here $p=p_{1,2}(\rho)$ is the pressure as a function of density in each layer,
with $p_{2}(\rho)-p_{1}(\rho)\ll[p_{2}(\rho)+p_{1}(\rho)]/2$.
In the equilibrium state the velocity potential $\varphi=0$, the enthalpy
$w_{1,2}(\bar\rho_{1,2}(z))=const_{1,2}-gz$, and the pressure is related to
the density by the hydrostatic formula
\begin{equation}
\bar p_{1,2}(z)=p_0-g\int_{h}^{z} \bar\rho_{1,2}(z) dz.
\end{equation}
Let us consider slow flows when $p_{1,2}=\bar p_{1,2}(z)+\tilde p_{1,2}$ and
$w_{1,2}\approx const_{1,2}-gz +\tilde p_{1,2}/\bar\rho(z)$, 
where $\tilde p_{1,2}$ are relatively small corrections to the pressure field due to
fluid flow. The equations of slow motion in the main order in $v/c$ take the form
\begin{eqnarray}\label{Bern_appr}
\partial_t\varphi_{1,2}+\frac{(\nabla\varphi_{1,2})^2}{2}
+\frac{\tilde p_{1,2}}{\bar\rho(z)}&=&0,\\
\label{varphi_equation}
\nabla\cdot(\bar\rho(z)\nabla\varphi_{1,2})&=&0.
\end{eqnarray}
It is the neglect of time derivative $\partial_t\rho$ in the continuity equation 
that allows us to exclude from the consideration acoustic degrees of freedom and
retain only ``soft''  modes as the internal waves which are conditioned by the
relatively small difference of the two equilibrium density profiles.
Compressibility of the medium in this model is manifested in form that
a volume of each fluid element at slow motion is effectively ``adapted'' to the 
equilibrium density $\bar\rho(z)$, expanding when going up and compressing 
when going down [since $\bar\rho'(z)<0$].

Let the shape of disturbed interface be given by equation $z=\eta({\bf x},t)$,
where ${\bf x}=(x,y)$ is the radius-vector in the horizontal plane, and let
the boundary values of the velocity potentials be 
$\psi_{1,2}({\bf x},t)=\varphi_{1,2}({\bf x},\eta(x,y,t),t)$.
At the free interface, the normal component $V_n$ of the velocity field
should be continuous, as well as the pressure. It is also clear that a local 
speed of boundary motion in the normal direction [for definiteness,
the normal vector ${\bf n}$ is directed from the first layer to the second one]
is equal to $V_n$. From these considerations, two kinematic conditions and one
dynamic condition are derived, which determine evolution of the system:
\begin{eqnarray}\label{V_n_contin}
&&\frac{\partial\varphi_1}{\partial n}\Big|_{z=\eta}
=\frac{\partial\varphi_2}{\partial n}\Big|_{z=\eta}\equiv V_n,\\
\label{kinematic_main}
&&\eta_t=V_n\sqrt{1+(\nabla\eta)^2},\\
&&\Big\{\bar\rho[\varphi_{1,t}-\varphi_{2,t}]
+\frac{\bar\rho}{2}[(\nabla\varphi_1)^2-(\nabla\varphi_2)^2]\Big\}\Big|_{z=\eta}
\nonumber\\
&&\qquad+g\int_{h}^{\eta} [\bar\rho_1(z)-\bar\rho_2(z) ]dz=0.
\label{dynamic_cond}
\end{eqnarray}

It follows form Eq.(\ref{V_n_contin}) that $\psi_1$ and $\psi_2$ are related to
each other by a linear integral dependence. Therefore, if we fix the difference
$\psi({\bf x},t)\equiv \psi_1-\psi_2$, then each potential will be fully 
determined. Taking into account the equalities
\begin{equation}\label{psi_t_phi_t}
\partial\psi_{1,2}/\partial t=[\partial\varphi_{1,2}/\partial t
+(\partial\varphi_{1,2}/\partial z)\eta_t]\Big|_{z=\eta},
\end{equation}
it is easy to check that the equations of motion for the two main functions
$\eta({\bf x},t)$ and $\psi({\bf x},t)$ possess the Hamiltonian structure
\begin{equation}\label{Hamiltonian_equations}
\bar\rho(\eta)\eta_t=\frac{\delta{\cal H}}{\delta\psi},\qquad 
-\bar\rho(\eta)\psi_t=\frac{\delta{\cal H}}{\delta\eta},
\end{equation}
with the corresponding Lagrangian
\begin{equation}
{\cal L}=\int \bar\rho(\eta)\eta_t\psi \,d^2{\bf x} -{\cal H}\{\eta,\psi\}.
\end{equation}
The Hamiltonian functional ${\cal H}\{\eta,\psi\}$ is given by the following
expression:
\begin{eqnarray}
{\cal H}&=&\int d^2{\bf x}\int_{0}^{\eta({\bf x})}\bar\rho(z)
\frac{(\nabla\varphi_1)^2}{2} dz\nonumber\\
&+&\int d^2{\bf x}\int_{\eta({\bf x})}^{+\infty}\bar\rho(z)
\frac{(\nabla\varphi_2)^2}{2} dz +g\int W(\eta)d^2{\bf x}\nonumber\\
&=&\frac{1}{2}\int\bar\rho(\eta) \psi V_nd^2{\bf x}+g\int W(\eta)d^2{\bf x},
\end{eqnarray}
where 
\begin{equation}
W'(\eta)=\int_h^\eta[\bar\rho_1(z)-\bar\rho_2(z)]dz,
\end{equation}
that is the Hamiltonian ${\cal H}$ is the sum of the kinetic energy and
an effective potential energy. Let us prove the above statements.

Indeed, the variation $\delta\psi$ entails some variations
$\delta\varphi_{1,2}$, and consequently --- a variation of the kinetic energy.
The corresponding variation of the Hamiltonian after integration by parts
is determined by a surface integral along the interface  $z=\eta({\bf x})$,
and it takes the form
\begin{eqnarray}
\delta{\cal H}\Big|_{\delta\psi}&=&\int_S\bar\rho(\nabla\varphi_1\cdot
{\bf n})\delta\psi_1 d S-\int_S\bar\rho(\nabla\varphi_2\cdot
{\bf n})\delta\psi_2 d S\nonumber\\
&=&\int \bar\rho(\eta) V_n\sqrt{1+(\nabla\eta)^2}\delta\psi d^2{\bf x}.
\end{eqnarray}
From here we have 
$\delta{\cal H}/\delta\psi=\bar\rho(\eta)V_n\sqrt{1+(\nabla\eta)^2}$ and,
making comparison with Eq.(\ref{kinematic_main}), we prove the first equation 
from Eqs.(\ref{Hamiltonian_equations}). Calculation of variational derivative
$\delta{\cal H}/\delta\eta$ is slightly more complicated, because when the
integration domain is varied, we have to ensure that after the interface variation
the difference $\psi_1-\psi_2$ takes at the new boundary the same value 
$\psi({\bf x})$ which was before the variation at the old boundary.
It is easy to understand that due to the above requirement the values of 
the potentials at the place of the old boundary are changed after variation
$\delta\eta$ by small quantities
$\delta\psi_{1,2}^{{old}}=-\delta\eta(\partial_z\varphi_{1,2})|_{z=\eta}$.
Accordingly, variation of the kinetic energy in this case consists of two
contributions. The first contribution comes from the change of integration domain:
\begin{equation}
\delta{\cal K}^{(1)}\Big|_{\delta\eta}=\int
\frac{\bar\rho}{2}[(\nabla\varphi_1)^2-(\nabla\varphi_2)^2]\Big|_{z=\eta}
\delta\eta d^2{\bf x}.
\end{equation}
The second contribution is related to the changes of the potentials
$\varphi_{1,2}$  in non-varied domains due to variations of their 
boundary values by the quantities $\delta\psi_{1,2}^{{old}}$. It is easy to 
understand that this contribution is equal to
\begin{eqnarray}
&&\delta{\cal K}^{(2)}\Big|_{\delta\eta}=\int
(\delta{\cal H}/\delta\psi)(\delta\psi_{1}^{old}-\delta\psi_{2}^{old}) d^2{\bf x}
\nonumber\\
&=&\int \bar\rho V_n\sqrt{1+(\nabla\eta)^2}
[\partial_z\varphi_{2}-\partial_z\varphi_{1}]\Big|_{z=\eta}\delta\eta d^2{\bf x}.
\end{eqnarray}
Taking into account also variation of the effective potential energy, we obtain
as the result
\begin{eqnarray}
\frac{\delta{\cal H}}{\delta\eta}&=&
\frac{\bar\rho}{2}[(\nabla\varphi_1)^2-(\nabla\varphi_2)^2]\Big|_{z=\eta}
+g\int_{h}^{\eta} [\bar\rho_1(z)-\bar\rho_2(z)]dz
\nonumber\\
&-&\bar\rho V_n\sqrt{1+(\nabla\eta)^2}
[\partial_z\varphi_{1}-\partial_z\varphi_{2}]\Big|_{z=\eta}. 
\end{eqnarray}
Looking at Eqs.(\ref{dynamic_cond}) and (\ref{psi_t_phi_t}), we obtain from here
the second equation of the Eqs.(\ref{Hamiltonian_equations}).

The Hamiltonian nature of the system under consideration in principle allows us 
to apply to it the standard set of methods \cite{ZK97}. However, a technical
difficulty is that the kinetic energy is not expressed directly but through
solutions of the partial derivative equation (\ref {varphi_equation}) 
with non-constant coefficients, and in domains with a curved boundary
$z=\eta({\bf x})$. Let us nevertheless suppose that particular solutions
of Eq.(\ref{varphi_equation}) are known in the form of linear combinations
\begin{equation}\label{phi_k}
\varphi_{\bf k}({\bf x},z)=[A\Phi^{(-)}_k(z)+B\Phi^{(+)}_k(z)]
e^{i{\bf k}\cdot{\bf x}},
\end{equation}
with decaying at $z\to+\infty$ functions $\Phi^{(-)}_k(z)$, and with
growing at $z\to+\infty$ functions $\Phi^{(+)}_k(z)$.
In other words, for every $k$ a general solution is known for the following
equation,
\begin{equation}
\Phi''(z)+\frac{\bar\rho'(z)}{\bar\rho(z)}\Phi'(z)-k^2\Phi(z)=0.
\end{equation}
Then for approximate calculation of the Hamiltonian at small deviations
$\zeta({\bf x},t)=\eta({\bf x},t)-h$, with the condition $|\nabla\zeta|\ll 1$,
one can write
\begin{eqnarray}
\!&&\!\varphi_1\!({\bf x},z)\!=\!\!\int\!\!\frac{d^2{\bf k}}{(2\pi)^2}
[A^{(1)}_{\bf k}\!\Phi^{(-)}_k\!(z)\!+\!B_{\bf k}^{(1)}\!\Phi^{(+)}_k\!(z)]
e^{i{\bf k}\cdot{\bf x}},
\\
\!&&\!\varphi_2({\bf x},z)=\int\frac{d^2{\bf k}}{(2\pi)^2}
A_{\bf k}^{(2)}\Phi^{(-)}_k(z) e^{i{\bf k}\cdot{\bf x}}.
\end{eqnarray}
After that from the set of boundary conditions
1) $\partial_z\varphi_1({\bf x},0)=0$; 
2) $\partial_n\varphi_1({\bf x},h+\zeta({\bf x}))=
\partial_n\varphi_2({\bf x},h+\zeta({\bf x}))$; and
3) $\varphi_1({\bf x},h+\zeta({\bf x}))-
\varphi_2({\bf x},h+\zeta({\bf x}))=\psi({\bf x})$ it is possible to find the
unknown function $A^{(1)}_{\bf k}$, $B^{(1)}_{\bf k}$, and  $A^{(2)}_{\bf k}$
[and consequently the required quantity $V_n({\bf x})$] in the form of an 
expansion in $\zeta$. Such a method of presenting the Hamiltonian as
a series in the small parameter of characteristic wave steepness
is generally used in the theory of surface water waves \cite{Z68,Z99,ZK97}. 
In particular, this method allows us to obtain the dispersion relation for
low-amplitude internal waves: 
\begin{equation}\label{omega_k_general}
\omega_k^2=\tilde g(h)\frac{D_1(h,k) D_2(h,k)}{[D_2(h,k)+D_1(h,k)]}, 
\end{equation}
where $\tilde g(h)$  is a renormalized gravity acceleration:  
$\tilde g(h)=g[\bar\rho_1(h)-\bar\rho_2(h)]/\bar\rho(h)$, and the 
short-hand notations have been used:
\begin{eqnarray}
D_1(h,k)&=&\frac{\Phi'^{(+)}_k(h)\Phi'^{(-)}_k(0)-\Phi'^{(-)}_k(h)\Phi'^{(+)}_k(0)}
{\Phi^{(+)}_k(h) \Phi'^{(-)}_k(0)- \Phi^{(-)}_k(h) \Phi'^{(+)}_k(0)},\\
D_2(h,k)&=&-\frac{ \Phi'^{(-)}_k(h)}{ \Phi^{(-)}_k(h)}.
\end{eqnarray}
Note that $D_1(h,k)>0$ and $D_2(h,k)>0$.

As to the system under consideration, here in some cases another way can be
suitable how to calculate the Hamiltonian. Since the kinetic energy takes the form
$$
{\cal K}=\frac{1}{2}\int ({\bf j}\cdot{\bf v})d^2{\bf x} dz, 
$$
where ${\bf j}=\bar\rho{\bf v}$ is the divergence-free field of the current 
density, we can introduce for ${\bf j}$ a vector potential ${\bf A}$ which
satisfies the equation 
\begin{equation}\label{A_equation}
\mbox{curl}\,\frac{1}{\bar\rho(z)}\,\mbox{curl}\,{\bf A}=
{\bf \Omega}\equiv\mbox{curl}\,{\bf v},
\end{equation}
with the boundary condition 
$[\partial_xA^{(y)}(x,y,0)-\partial_yA^{(x)}(x,y,0)]=0$. After that the kinetic 
energy can be re-written as follows,
\begin{eqnarray}
{\cal K}&=&\frac{1}{2}\int {\bf A}\cdot {\bf \Omega}\,\,d^2{\bf x} dz\nonumber\\
&=&\frac{1}{2}\int G_{ik}({\bf r}_1,{\bf r}_2)
\Omega_i({\bf r}_1)\Omega_k({\bf r}_2)\,\,d^3{\bf r}_1 d^3{\bf r}_2,
\end{eqnarray}
where $G_{ik}({\bf r}_1,{\bf r}_2)$ is the Green's function for 
Eq.(\ref{A_equation}). As far as the (singular) vorticity field ${\bf \Omega}$
is totally concentrated at the interface $z=\eta({\bf x})$, and the vortex lines
coincide with levels of the function $\psi({\bf x})$ at that surface,
the half-space integration will reduce to integration along the surface
$z=\eta({\bf x})$ by means of the change
\begin{equation}\label{Omega_to_psi}
(\Omega^{(x)},\Omega^{(y)},\Omega^{(z)})d^3{\bf r}\to
(\psi_y, -\psi_x, \psi_y\eta_x-\psi_x\eta_y)dxdy.
\end{equation}

As the simplest example, in this work an exponential profile
$\bar\rho(z)=\rho_0\exp(-2\alpha z)$ of the equilibrium density will be
considered, when Eq.(\ref{A_equation}) after substitution
${\bf A}=\rho_0 e^{-2\alpha z}{\bf F}$ turns into an equation
with constant coefficients. Generally speaking, if taken globally,
such a dependence contradicts to adiabatic equations of state for real gases, 
for those we rather have $p\approx C_1 \rho^\gamma$, where
$\gamma$ is the adiabatic exponent [for single-atom gases $\gamma=5/3$,
for gases consisting of two-atom molecules $\gamma=7/5$], and therefore
 $\bar\rho(z)\approx C_2(z_0-z)^{1/(\gamma-1)}$, where $z_0$ is the altitude
of the upper edge of the atmosphere. Nevertheless, locally on the vertical
coordinate near $z=h$, every realistic dependence $\bar\rho(z)$ is approximated
by an exponent, provided not very long waves are considered.
We still would like to note that the case 
$\bar\rho(z)\approx C_2(z_0-z)^{1/(\gamma-1)}$
also admits analytic investigation, though more difficult, since the functions
$\Phi_k^{(\pm)}(z)$ in the particular solutions (\ref{phi_k}) of
Eq.(\ref{varphi_equation}) are expressed in that case through the modified
Bessel functions Бесселя $I_\nu$ and $K_\nu$, with the index
$\nu=[(\gamma-1)^{-1}-1]/2$:
\begin{eqnarray}
\Phi^{(-)}_k(z)&=&[k(z_0-z)]^{-\nu}I_\nu(k(z_0-z)),\\
\Phi^{(+)}_k(z)&=&[k(z_0-z)]^{-\nu}K_\nu(k(z_0-z)).
\end{eqnarray}

\section{The case of exponential profile of equilibrium density}

Thus, we have to find the Hamiltonian of our system in an explicit form for
$\bar\rho(z)=\rho_0\exp(-2\alpha z)$, and at the beginning we will solve 
Eq.(\ref{A_equation}). Consider here simpler case $\alpha h\gg 1$, when
the presence of the flat lower boundary at при $z=0$ is not important,
because the corresponding contribution will be shown later to be of the order
 $\exp(-2h\alpha)$. More cumbersome 3D solution for the vector potential
in the presence of the boundary  $z=0$ is given in the Appendix.
To solve Eq.(\ref{A_equation}), we use the substitution 
${\bf A}=\rho_0 e^{-2\alpha z}{\bf F}$ and re-write the equation in 
Fourier representation:
$i{\bf k}\times[(i{\bf k}-2\alpha{\bf e}_z)\times {\bf F}_{\bf k}]
={\bf\Omega}_{\bf k}$.
Applying the well-known formula for the double vector cross-product
and choosing the gauge  $({\bf k}\cdot{\bf F}_{\bf k})=0$, we immediately
arrive at a simple equation
\begin{equation}
\label{F_equation}
[k^2+2i\alpha({\bf k}\cdot{\bf e}_z)]{\bf F}_{\bf k}={\bf\Omega}_{\bf k}.
\end{equation}
Now we write down the decaying at the infinity solution of the above equation:
\begin{eqnarray}\label{F_particular}
{\bf F}({\bf r})&=&\int\frac{d^3{\bf k}}{(2\pi)^3}
\frac{{\bf\Omega}_{\bf k}e^{i{\bf k}\cdot{\bf r}}}
{[k^2+2i\alpha({\bf k}\cdot{\bf e}_z)]}\nonumber\\
&=&\int\frac{\exp[\alpha (z-z_1-|{\bf r}-{\bf r}_1|)]}{4\pi|{\bf r}-{\bf r}_1|}
{\bf\Omega}({\bf r}_1)d^3{\bf r}_1.
\end{eqnarray}
Accordingly, the kinetic energy of the 3D system, without taking into account
the flat rigid boundary, is given by the following expression:
\begin{equation}
{\cal K}=\frac{\rho_0}{8\pi}\int\frac{e^{-\alpha|{\bf r}_2-{\bf r}_1|}}
{|{\bf r}_2-{\bf r}_1|}e^{-\alpha (z_2+z_1)}
{\bf\Omega}({\bf r}_2)\cdot {\bf\Omega}({\bf r}_1)d^3{\bf r}_1d^3{\bf r}_2.
\end{equation}
Passing with the help of formula (\ref{Omega_to_psi}) from the space integration
to the surface integration where singular vorticity field is distributed, we
arrive at the expression in terms of $\eta$ and $\psi$,
\begin{eqnarray}
{\cal K}&\!=\!&\frac{\rho_0}{8\pi}\!
\int \!\frac{\exp[-\alpha\sqrt{|{\bf x}_1\!-\!{\bf x}_2|^2+(\eta_1\!-\!\eta_2)^2}-
\alpha(\eta_1\!+\!\eta_2)]}
{\sqrt{|{\bf x}_1-{\bf x}_2|^2+(\eta_1-\eta_2)^2}}\nonumber\\
&&\qquad \times\{\nabla\psi_1\cdot\nabla\psi_2+
[\nabla\psi_1\times\nabla\eta_1]\cdot[\nabla\psi_2\times\nabla\eta_2]\}
\nonumber\\
&&\qquad\qquad \times d^2{\bf x}_1d^2{\bf x}_2 ,
\end{eqnarray}
where $\nabla\eta$ and $\nabla\psi$  are 2D gradients. If necessary,
a weakly nonlinear regime in the wave dynamics can be easy considered through 
expansion of the above expression in powers of $\psi$ and $\zeta$.

Let us now turn our attention to planar flows. Note that in 2D case
$F_{\bf k}$ and $\Omega_{\bf k}$ are in the essence (pseudo) scalar quantities.
The presence of the boundary at $z=0$ can be taken into account by a variant
of the ``image method'', and as the result we have 
\begin{eqnarray}
F(x,z)&=&\frac{1}{2\pi}\int 
\Big[K_0(\alpha\sqrt{(x-x_1)^2+(z-z_1)^2})\nonumber\\
&&\qquad-K_0(\alpha\sqrt{(x-x_1)^2+(z+z_1)^2})\Big]\nonumber\\
&&\qquad\times e^{\alpha (z-z_1)}\Omega(x_1,z_1)dx_1dz_1,
\end{eqnarray}
where $K_0(r)$ is the well-known Macdonald function. We provide below two
of many possible integral representations for this function:
\begin{eqnarray}
&&K_0(\sqrt{a^2+b^2})=\int\frac
{d^2{\bf k}}{2\pi}\frac{e^{ik_1a+ik_2b}}{k_1^2+k_2^2+1}\nonumber\\
&&=\int_{-\infty}^{+\infty}
\frac{\exp(ika -|b|\sqrt{k^2+1})}{2\sqrt{k^2 +1}}dk.
\label{K_0}
\end{eqnarray}
Consequently, the Green's function in this case takes the form
\begin{eqnarray}
&&G(x_1,x_2,z_1,z_2)=\frac{\rho_0}{2\pi}\Big[
K_0\left(\alpha\sqrt{(x_1\!-\!x_2)^2+(z_1\!-\!z_2)^2}\right)\nonumber\\
&&\quad-K_0\left(\alpha\sqrt{(x_1-x_2)^2+(z_1+z_2)^2}\right)
\Big]e^{-\alpha(z_1+z_2)}.
\label{G_2D_exp}
\end{eqnarray}
The expression for the kinetic energy of the two-layer flow looks as follows:
\begin{eqnarray}
{\cal K}_{2D}&=&\frac{\rho_0}{4\pi}\int \Big[
K_0\left(\alpha\sqrt{(x_1-x_2)^2+(\eta_1-\eta_2)^2}\right)\nonumber\\
&&\qquad-K_0\left(\alpha\sqrt{(x_1-x_2)^2+(\eta_1+\eta_2)^2}\right)\Big]\nonumber\\
&&\qquad\times e^{-\alpha(\eta_1+\eta_2)}\psi'_1\psi'_2 dx_1 dx_2,
\end{eqnarray}
where $\psi'=\partial\psi/\partial x$. Use of formulas (\ref{K_0}) 
allows us to represent this functional in a slightly different form:
\begin{eqnarray}\label{K2D_representation}
{\cal K}_{2D}&=&\frac{\rho_0}{2}\int dx_1 dx_2\psi'_1\psi'_2
e^{-\alpha(\eta_1+\eta_2)}e^{ik(x_1-x_2)}\nonumber\\
&\times&\!\!\!\int\!\!\frac{[e^{-|\eta_1-\eta_2|\sqrt{k^2+\alpha^2}}-
e^{-(\eta_1+\eta_2)\sqrt{k^2+\alpha^2}}]}
{2\sqrt{k^2+\alpha^2}}\frac{dk}{2\pi}.
\end{eqnarray}
As it will be shown later, such a representation is suitable for consideration
of long-wave asymptotics in the nonlinear wave dynamics. Besides that, it also
allows us to find easily the dispersion relation for linear waves.
Indeed, from Eq.(\ref{K2D_representation}) it is obvious that in the quadratic
approximation the Hamiltonian is given by the formula
\begin{eqnarray}\label{H_quadratic}
{\cal H}^{[2]}_{2D}&=&\frac{\rho_0e^{-2\alpha h}}{2}\int
\Big[\frac{[1-e^{-2h\sqrt{k^2+\alpha^2}}]}{2\sqrt{k^2+\alpha^2}}
k^2 \psi_{-k}\psi_k \nonumber\\
&&\qquad \qquad+\tilde g(h)\zeta_{-k}\zeta_k\Big]\frac{dk}{2\pi},
\end{eqnarray}
Solving the corresponding linearized equations of motion for the Fourier components
 $\zeta_k(t)$ and $\psi_k(t)$, we find quite nontrivial expression for
the dispersion relation:
\begin{equation}
\omega_k^2=\tilde g(h) k^2\frac{[1-e^{-2h\sqrt{k^2+\alpha^2}}]}{2\sqrt{k^2+\alpha^2}}.
\end{equation}
Note, the same dispersion law takes place in the 3D case, due to the isotropy
of the system in the horizontal plane [it is also confirmed by the formula
(\ref{omega_k_general})].

Now we consider the limiting case $\alpha\eta\ll 1$ and typical wave numbers $k$ 
satisfying the conditions $\alpha\eta\lesssim k\eta\ll 1$. Expanding the exponents
in integral (\ref{K2D_representation}) in powers of the small arguments, we
obtain an approximate kinetic energy functional up to the first order in
$\alpha\eta$,
\begin{eqnarray}\label{K2D_approx_eta_psi}
{\cal K}_*\{\eta,\psi\}&=&\frac{\rho_0}{2}
\int \eta(1-2\alpha \eta) (\psi')^2dx\nonumber\\
&-&\frac{\rho_0}{2}\int  (\psi'\eta)
[-\hat\partial_x^2+\alpha^2]^{1/2}(\psi'\eta )\, dx.
\end{eqnarray}
Let us introduce a new unknown function $q=[1-\exp(-2\alpha\eta)]/(2\alpha)$,
which up to the constant factor $\rho_0$ is the canonically conjugate for
function $\psi$, and then re-write the approximate Hamiltonian in terms of
$q$ and $\psi$:
\begin{eqnarray}\label{H2D_approx_q_psi}
{\cal H}_*\{q,\psi\}&=&\frac{\rho_0}{2}\int q(1-\alpha q) (\psi')^2dx\nonumber\\
&-&\frac{\rho_0}{2}\int  (\psi' q)
[-\hat\partial_x^2+\alpha^2]^{1/2}(\psi'q ) dx\nonumber\\
&+&\rho_0 \tilde g(0)\int\left[\frac{q^2}{2}+\alpha\beta\frac{q^3}{3}\right]dx,
\end{eqnarray}
where $\beta$  is a dimensionless parameter depending on behavior of the difference
$[\bar\rho_1(z)-\bar\rho_2(z)]$ near $z=0$. Considering propagation of relatively
small but finite disturbances $\tilde q(x,t)= q(x,t)-\bar q$, it is possible by a
standard procedure to derive weakly nonlinear equation for $u(x,t)=\psi_x$, which
describes a slow evolution of uni-directional wave under the influence of weak
dispersion:
\begin{equation}\label{slow_evolution}
u_t+\bar c u_x+ \bar a uu_x-
\frac{\bar c \bar q}{2}\{[-\hat\partial_x^2+\alpha^2]^{1/2}-\alpha\}u_x=0,
\end{equation}
where the speed of long linear waves is
$\bar c\approx[\tilde g(0)\bar q]^{1/2}$, and the coefficient 
$\bar a\approx 3/2$. Equation of such kind is called sometimes ``Smith's equation''
after the  work by Ronald Smith \cite{Smith_1972} where it arose for the first time 
in context of continental-shelf oceanic waves. 
It is interesting to note that the special form
of the dispersive term makes the above equation intermediate
between the two famous integrable models, namely the Korteweg-de Vries equation
and the Benjamin-Ono equation \cite{BENJAMIN,ONO}. In this sense 
the Smith's equation is similar to the Intermediate Long Wave equation
(ILW) (see, e.g., \cite{Joseph,Chen_Lee_1979,Choi_Camassa_1999,CGK_2005,deZarate_Nachbin}),
but contrary to ILW the Smith's equation is not integrable, 
as it was established in Ref.\cite{ABFS_1989}.

\section{Planar flows with piecewise constant potential vorticity}

Now we would like to make an important generalization of the Hamiltonian theory
which is possible for 2D isentropic flows [in $(x,z)$ plane], namely we will
take into account the fact that potential vorticity
$\tilde\gamma=-\Omega^{(y)}/\rho$ in the 2D case is governed by the advection 
equation
\begin{equation}
\tilde\gamma_t+{\bf v}\cdot\nabla \tilde\gamma=0.
\end{equation}
This conservation law for the potential vorticity along each fluid particle trajectory 
allows us at consideration of planar flows with a piecewise constant
function $\tilde\gamma(x,z,t)$ to follow only the motion of boundaries where
$\tilde\gamma$ is discontinuous. In the present paper it is assumed for simplicity
that $\tilde\gamma$ has a single jump, and this jump coincides with the interface
between the layers  $z=\eta(x,t)$, but generally this coincidence is not necessary
and a separate curve $z=\eta_*(x,z,t)$ can be considered where the jump takes
place.

Let (sufficiently small) potential vorticities in the layers be $\gamma_{1,2}$, 
so that the corresponding stationary shear flows $U_{1,2}(z)\ll c$ satisfy the
conditions (we neglect the difference between $\bar\rho_{1,2}$ and $\bar\rho$)
\begin{equation}
-U'_{1,2}(z)=\gamma_{1,2}\bar\rho(z).
\end{equation}
We shall suppose that in the stationary state the velocity profile has a ``break''
at $z=h$, that is $U_{1,2}(z)=-\gamma_{1,2}\mu(z)$, where
\begin{equation}\label{mu_def}
\mu(z)=\int_h^z\bar\rho(\xi) d\xi.
\end{equation}
A 2D velocity field in each layer now takes the form
\begin{equation}
{\bf v}_{1,2}(x,z,t)=(U_{1,2}(z)+\partial_x\varphi_{1,2}(x,z,t), \quad
\partial_z\varphi_{1,2}(x,z,t)),
\end{equation}
with the potentials $\varphi_{1,2}$ satisfying the same equation
(\ref{varphi_equation}): 
$\nabla\cdot\bar\rho\nabla\varphi_{1,2} =0$, and it implies the existence of the
corresponding stream functions $\vartheta_{1,2}(x,z,t)$:
\begin{equation}
\bar\rho\partial_x\varphi_{1,2}=\partial_z\vartheta_{1,2},\qquad
\bar\rho\partial_z\varphi_{1,2}=-\partial_x\vartheta_{1,2}.
\end{equation}
Instead of Eq.(\ref{Bern_appr}), we have to deal now with its generalization:
\begin{equation}\label{Bern_appr_generalized}
\partial_t\varphi_{1,2}+\gamma_{1,2}\vartheta_{1,2}
+U_{1,2}(z)\partial_x\varphi_{1,2}+\frac{(\nabla\varphi_{1,2})^2}{2}
+\frac{\tilde p_{1,2}}{\bar\rho(z)}=0,
\end{equation}
which regards the 2D Euler equation in the case of constant potential vorticity
under the condition $\nabla\cdot(\bar\rho{\bf v})=0$. 
Taking into account that the full stream functions of the flows under
consideration are
\begin{equation}
\Theta_{1,2}(x,z,t)=\vartheta_{1,2}(x,z,t)-U^2_{1,2}(z)/(2\gamma_{1,2}),
\end{equation}
equation (\ref{Bern_appr_generalized}) can be also represented as follows,
\begin{equation}\label{Bern_appr_generalized_2}
\partial_t\varphi_{1,2}+\gamma_{1,2}\Theta_{1,2}
+\frac{({\bf v}_{1,2})^2}{2}
+\frac{\tilde p_{1,2}}{\bar\rho(z)}=0.
\end{equation}
Now we note that at the interface $z=\eta(x,t)$ there are the equalities
\begin{eqnarray}
&&-\partial_x\Theta_1(x,\eta(x))=-\partial_x\Theta_2(x,\eta(x))
=\bar\rho(\eta)\eta_t\nonumber\\
&&=\bar\rho (\eta)V_n\sqrt{1+\eta'^2},
\end{eqnarray}
where $V_n=({\bf v}_1\cdot{\bf n})=({\bf v}_2\cdot{\bf n})$.

Demanding the pressure field to be continuous at $z=\eta(x,t)$ and
reasoning analogously to the case $\gamma_{1,2}=0$, we conclude that
the evolution equations for the 2D system possess the following structure,
\begin{eqnarray}\label{Ham_eq_gamma_eta}
\bar\rho(\eta)\eta_t&=&{\delta{\cal H}}/{\delta\psi},\\
\label{Ham_eq_gamma_psi}
-\bar\rho(\eta)\psi_t+\gamma\bar\rho(\eta)
\partial_x^{-1}[\bar\rho(\eta)\eta_t]&=&{\delta{\cal H}}/{\delta\eta},
\end{eqnarray}
where $\gamma=(\gamma_1-\gamma_2)$, and the Hamiltonian ${\cal H}$ is equal to the
sum of total kinetic energy and the effective potential energy.
By a direct calculation it is easy to check that the corresponding Lagrangian
for the above equations is
\begin{equation}\label{L_2D_gamma}
{\cal L}=\int \psi \mu_t \,d x +
\frac{\gamma}{2}\int \mu\partial_x^{-1}\mu_t\,d x -{\cal H}\{\mu,\psi\},
\end{equation}
where  $\mu=\mu(\eta)$ [see Eq.(\ref{mu_def})].  For internal waves in an
incompressible liquid, an analogous structure was obtained in
Refs.\cite{Goncharov,R_Y_2007}, with the difference that in our case $\mu(\eta)$
is a nonlinear function (see also Ref.\cite{Wahlen} about waves at the free surface
of a 2D incompressible fluid with a constant vorticity).

It is interesting to note that in the quadratic approximation the Lagrangian
(\ref {L_2D_gamma}) take the form
\begin{equation}
{\cal L}^{[2]}=\bar\rho(h)\int \psi \zeta_t \,d x 
+\frac{\gamma \bar\rho^2(h)}{2}\int \zeta\partial_x^{-1}\zeta_t\,d x 
-{\cal H}^{[2]}\{\zeta,\psi\}.
\end{equation}
Moreover, it is easy to show that the functional ${\cal H}^{[2]}\{\zeta,\psi\}$ 
does not depend on $\gamma_1$ and $\gamma_2$ [dependence on $\gamma_1$ and  $\gamma_2$ 
appears only in higher orders]:
\begin{equation}\label{H_quadratic_general}
{\cal H}^{[2]}=\frac{\bar\rho(h)}{2}\int\left[
N(h,k) k^2\psi_{-k}\psi_k +\tilde g(h)\zeta_{-k}\zeta_k\right]\frac{dk}{2\pi}.
\end{equation}
Function $N(h,k)$ is expressed through the Green's function $G[(x_2-x_1),z_1,z_2]$
by the following formula:
\begin{equation}
\bar\rho(h)N(h,k)=\int_{-\infty}^{+\infty} G[x,h,h]e^{-ikx}dx.
\end{equation}
It should be noted that $\omega^2_0(k)=\tilde g(h)k^2N(h,k)$ is the dispersion law
in the case $\gamma_1=\gamma_2=0$ [compare with (\ref{omega_k_general})].
For example, with the exponential profile of the equilibrium density
the quadratic Hamiltonian is given by expression (\ref {H_quadratic}). Solving
the corresponding linear equations,
\begin{eqnarray}
\dot\zeta_k&=&N(h,k)k^2\psi_k,
\\
-\dot\psi_k+\gamma \bar\rho(h)\frac{\dot\zeta_k}{ik}&=&\tilde g(h)\zeta_k,
\end{eqnarray}
we obtain the dispersion law for linear waves at $\gamma\not = 0$:
\begin{eqnarray}\label{omega_k_gamma}
\omega_k&=&\frac{1}{2}\gamma\bar\rho(h)kN(h,k)\nonumber\\
&+&\sqrt{\left[\gamma\bar\rho(h)kN(h,k)\right]^2/4
+\tilde g(h)k^2N(h,k)}.
\end{eqnarray}

Since the singular part of the vorticity field (concentrated at the interface) 
is determined by the relation
$-\Omega_s=[\psi'-\gamma\mu(\eta)]\delta(z-\eta(x,t))$, 
where  $\delta(z-\eta(x,t))$ is the Dirac's function, it is convenient to
introduce the new unknown variable,
\begin{equation}
p(x,t)=\psi-\gamma\partial_x^{-1}\mu.
\end{equation}
In variables $\{\mu,p\}$ the Lagrangian takes the form (the sign in front of the
second term has changed)
\begin{equation}
{\cal L}=\int p \mu_t \,d x -\frac{\gamma}{2}\int \mu\partial_x^{-1}\mu_t\,d x
-{\cal H}\{\mu,p\}.
\end{equation}
Now, besides the singular part of the vorticity, there is also a distributed
part, and the full vorticity field is given by the formula
\begin{eqnarray}\label{Omega_total}
-\Omega(x,z,t)&=&p'(x,t)\delta[z-\eta(x,t)]+\nonumber\\
 &+&\gamma_2\bar\rho(z)+\gamma\bar\rho(z)\theta[\eta(x,t)-z],
\end{eqnarray}
where $\theta[\eta(x,t)-z]$ is the unit jump function (the Heaviside's function).
The Hamiltonian of the 2D system is determined with the help of the Green's
function $G[(x_2-x_1),z_1,z_2]$ by the following expression:
\begin{eqnarray}
{\cal H}&=&g\int W(\eta)d x+\frac{1}{2}\int G[(x_2-x_1),z_1,z_2]\nonumber\\
&&\qquad\times 
\Omega(x_1,z_1)\Omega(x_2,z_2)dx_1dz_1 dx_2dz_2,
\end{eqnarray}
where Eq.(\ref{Omega_total}) should be substituted, and after the integrations
$\eta$ should be expressed through $\mu$. Let us remind that in the case
$\bar\rho(z)=\rho_0\exp(-2\alpha z)$ the Green's function is given by
Eq.(\ref{G_2D_exp}). Let us also note that in the absence of the density jump
a class of flows is possible with $p\equiv 0$. In that case the dynamics of
the vorticity waves is determined by the Lagrangian
${\cal L}_\gamma=-({\gamma}/{2})\int \mu\partial_x^{-1}\mu_t\,d x-
{\cal H}_\gamma\{\mu\}$, and the dispersion law for such waves is expressed by
the formula (\ref{omega_k_gamma}) where $\tilde g(h)=0$ should be put.

\section{Discussion}

In this work, a compressible two-layer atmospheric model has been suggested,
intended for theoretical study of internal waves at the interface between
two isentropic layers of a gas with nearly equal values of specific entropy.
In the derivation of the approximate equations it was supposed that the 
flow velocities are small compared with a local speed of sound. It should be noted
that this condition puts the lower limit for characteristic wave numbers:
$k\gtrsim|\bar\rho'(h)|/\bar\rho(h)$, because at longer scales the velocity 
field penetrates into the upper layer rather far, where in view of constant
entropy the temperature is small together with a local speed of sound, and it
violates the starting-point assumption of the model. To some extent
the above limitation is softened if somewhere above the second layer there is 
the third layer, with very high temperature, and therefore the boundary between the
second layer and the third layer can be effectively treated as a ``rigid lid''.
However one should remember that in the long-wave limit (in the Earth conditions
it corresponds to hundreds and thousands kilometers), nonuniform horizontal 
motions of the whole atmosphere become important. Those flows are approximately
described by a ``shallow water theory'' with adding the Coriolis force,
and they lead to variations of a quasi-equilibrium density profile.
Besides that, the Coriolis force violates the potentiality of the flow.
Thus, the suggested here theory can describe waves with lengths not longer than
a few kilometers

In the present work, only first steps have been made in the study of internal waves
in the atmosphere within the compressible two-layer model. Promising directions of
further research can be outlined as follows. First, a generalization of the model 
is evident for more layers and for continuous limit, which will enrich it because
an interaction between several interfaces in many cases is able to introduce new
interesting effects as instabilities etc. 
Second, we should mention a wide class of problems
about interaction of internal waves and mountains, which also can be studied
with the help of this model. Third, nonlinear wave dynamics can be simulated
numerically. Fourth, an analogous Hamiltonian formulation is possible for
consideration of axisymmetric flows with a piecewise constant generalized
potential vorticity. Fifth, it seems likely that analogous finite-layer models
are possible not only in the Eulerian hydrodynamics, but in a wider class
of conservative hydrodynamic systems as well, for instance, in the hydrodynamics
of a relativistic fluid placed in a strong static gravitational field described by
a metric 4-tensor. Accordingly, there is a perspective of application of a similar
theory to astrophysical problems, where the equilibrium density possesses 
the spherical symmetry, as a rule.

These investigations were supported by RFBR 
(grants 09-01-00631 and 07-01-92165),
by the ``Leading Scientific Schools of Russia'' grant 6885.2010.2,
and by the Program ``Fundamental Problems of Nonlinear Dynamics'' 
from the RAS Presidium.

\appendix
\section{Correction to 3D Green's function due to the flat boundary}

To satisfy the boundary condition
$[\partial_x F^{(y)}(x,y,0)-\partial_y F^{(x)}(x,y,0)]=0$, which ensures zero
normal component of the  velocity field at the rigid flat boundary, we add
to the particular solution (\ref{F_particular}) of the non-homogeneous equation
(\ref{F_equation}) some specially selected solution of the corresponding
homogeneous equation, decaying at $z\to+\infty$:
\begin{equation}\label{F_minus}
{\bf F}^{(-)}({\bf x},z)=\int\frac{d^2{\bf k}}{(2\pi)^2}{\bf f}_{\bf k}
\exp[i{\bf k}\cdot{\bf x}+z(\alpha-\sqrt{k^2+\alpha^2})],
\end{equation}
where ${\bf f}_{\bf k}=(f^{(x)}_{\bf k},f^{(y)}_{\bf k},0)$ 
satisfies the condition of 2D transversal gauge $({\bf k}\cdot{\bf f}_{\bf k})=0$.
It is not difficult to understand that ${\bf f}_{\bf k}$ should be taken in
the following form (here and later on $\kappa$ and $\nu$ are tensorial indices
in the horizontal plane):
\begin{equation}\label{f_k}
 f^{(\kappa)}_{\bf k}=-\left(\delta_{\kappa\nu}-\frac{k_\kappa k_\nu}{k^2}\right)
\int\frac{d\xi}{2\pi}\frac{\Omega^{(\nu)}({\bf k},\xi)}{(k^2+\xi^2+2i\alpha\xi)},
\end{equation}
where $\Omega^{(\nu)}({\bf k},\xi)\equiv
\int\Omega^{(\nu)}({\bf x}_1,z_1)e^{-i{\bf k}\cdot{\bf x}_1-i\xi z_1}
d^2{\bf x}_1dz_1$ is the Fourier image of the horizontal component of the 
vorticity field. Now we transform the integral  (\ref{f_k}):
\begin{eqnarray}
&&\int\frac{d\xi}{2\pi}\frac{\Omega^{(\nu)}({\bf x}_1,z_1)
e^{-i{\bf k}\cdot{\bf x}_1-i\xi z_1}}
{(k^2+\xi^2+2i\alpha\xi)} d^2{\bf x}_1dz_1\nonumber\\
&&=\int\frac{\Omega^{(\nu)}({\bf x}_1,z_1)
e^{-i{\bf k}\cdot{\bf x}_1- z_1(\alpha+\sqrt{k^2+\alpha^2})}}
{2\sqrt{k^2+\alpha^2}} d^2{\bf x}_1dz_1.
\end{eqnarray}
For $\xi$-integration we have used the fact that ${\bf\Omega}({\bf x},z)$ 
is non-zero only at $z>0$, and therefore the integration contour can be closed in
in the lower complex half-plane. Then we substitute the result into
Eq.(\ref{F_minus}) and see that the conditioned by the flat boundary correction 
$G^{(-)}_{\kappa\nu}({\bf x}_1,{\bf x}_2,z_1,z_2)$ to the Green' function
actually depends on the variables ${\bf x}={\bf x}_2-{\bf x}_1$ and $s=z_1+z_2$,
and it is expressed by the following formula: 
\begin{eqnarray}
&&G^{(-)}_{\kappa\nu}({\bf x},s)=\rho_0e^{-\alpha s}
\int\left(\frac{k_\kappa k_\nu}{k^2}-\delta_{\kappa\nu}\right)\nonumber\\
&&\qquad\qquad\qquad\times
\frac{e^{i{\bf k}\cdot{\bf x}- s\sqrt{k^2+\alpha^2}}}
{2\sqrt{k^2+\alpha^2}}\frac{d^2{\bf k}}{(2\pi)^2}\nonumber\\
&=&-\rho_0e^{-\alpha s}(\delta_{\kappa\nu}-
\partial_{\kappa}\partial_{\nu}\hat\Delta_{\bf x}^{-1})
\frac{\exp[-\alpha\sqrt{{\bf x}^2+s^2}]}{4\pi\sqrt{{\bf x}^2+s^2}},
\end{eqnarray}
where $\hat\Delta_{\bf x}^{-1}$ is the inverse 2D Laplace operator. Let us
introduce the notation  $D(|{\bf x}|,s) =
\Delta_{\bf x}^{-1}[\exp(-\alpha\sqrt{{\bf x}^2+s^2})/
\sqrt{{\bf x}^2+s^2}]$.
In virtue of the definition, function $D(r,s)$  satisfies the equation
\begin{equation}
\frac{1}{r}\frac{\partial}{\partial r}\left(r\frac{\partial D}{\partial r}\right)=
\frac{\exp(-\alpha\sqrt{r^2+s^2})}{\sqrt{r^2+s^2}},
\end{equation}
from which we obtain by a simple integration
\begin{equation}\label{rD_r}
r{\partial D}/{\partial r}=[\exp(-\alpha s)-\exp(-\alpha\sqrt{r^2+s^2})]/\alpha.
\end{equation}
It should be noted that the second derivatives
$\partial_{\kappa}\partial_{\nu}D(|{\bf x}|,s)$ can be expressed through the
combination $r^{-1}{\partial D}/{\partial r}$:
\begin{equation}
\partial_{\kappa}\partial_{\nu}D(|{\bf x}|,s)=\frac{x_\kappa x_\nu}{r}
\frac{\partial}{\partial r}\left(\frac{1}{r}\frac{\partial D}{\partial r}\right)
+\delta_{\kappa\nu}\frac{1}{r}\frac{\partial D}{\partial r}.
\end{equation}
Collecting the obtained expressions and taking into account Eq.(\ref{rD_r}),
we write the required correction to 3D Green's function in the final form:
\begin{eqnarray}
&&G^{(-)}_{\kappa\nu}({\bf x},s)=
\rho_0\left(\delta_{\kappa\nu}-2\frac{x_\kappa x_\nu}{{\bf x}^2}\right)\nonumber\\
&&\qquad\times
\frac{[\exp(-2\alpha s)-\exp(-\alpha s-\alpha\sqrt{{\bf x}^2+s^2})]}
{4\pi\alpha {\bf x}^2}\nonumber\\
&&+\rho_0\left(\frac{x_\kappa x_\nu}{{\bf x}^2}-\delta_{\kappa\nu}\right)
\frac{\exp(-\alpha s-\alpha\sqrt{{\bf x}^2+s^2})}{4\pi\sqrt{{\bf x}^2+s^2}}.
\end{eqnarray}

\end{document}